\documentstyle[prl,aps]{revtex}
\include{epsf} 
\begin{document}
\wideabs{
\title{$\bf O(\alpha)$ Radiative Correction to the Casimir 
Energy for Penetrable Mirrors}
\author{M.\ Bordag\thanks{email: Michael.Bordag@itp.uni-leipzig.de}}
\address{Universit\"at Leipzig, Institut f\"ur Theoretische Physik,
Augustusplatz 10, D-04109 Leipzig, Federal Republic of Germany}
\author{K.\ Scharnhorst\thanks{email: scharnh@physik.hu-berlin.de}}
\address{Humboldt-Universit\"at zu Berlin, Institut f\"ur Physik,
Invalidenstr.\ 110, D-10115 Berlin, Federal Republic of Germany}
\date{\today}
\maketitle
\begin{abstract}
The leading radiative correction to the Casimir energy for two
parallel penetrable mirrors is calculated within QED perturbation
theory. It is found to be of the order $\alpha$ like the known
radiative correction for ideally reflecting mirrors from which it
differs only by a monotonic, powerlike function of the frequency at
which the mirrors become transparent. This shows that the
$O(\alpha^2)$ radiative correction calculated recently by Kong and
Ravndal for ideally reflecting mirrors on the basis of effective field
theory methods remains subleading even for the physical case of
penetrable mirrors.
\end{abstract}
\pacs{12.20Ds}
}
The existence of zero point fluctuations of all quantum fields 
is shaping our modern view of the physical vacuum which is 
thought to be a complicated medium \cite{vacu}. 
As these vacuum fluctuations cannot be made to disappear 
completely, their modification 
by means of external fields or boundary conditions as they occur 
at conducting surfaces plays a significant role in studying the
vacuum properties. The Casimir effect \cite{casi1,casi2}, i.e.\ the 
mutual attraction of two parallel uncharged conducting plates 
(mirrors) in vacuo, represents a key phenomenon in the modern study 
of the vacuum (in recent experiments it has become well established 
quantitatively \cite{lamo,mohi}). For ideally reflecting
parallel mirrors the distance-dependent part of the vacuum energy
(per unit area of the mirrors) reads ($\hbar = c=1$)
\begin{equation}
\label{casienergy}
E_0\ =\ -\ \frac{\pi^2}{720}\ \frac{1}{d^3},
\end{equation}
where $d$ denotes the distance between the mirrors.
Within quantum electrodynamics (QED), 
this result can be calculated by means of free field theory
and the Casimir pressure $p$ can be derived from
it using the relation $p = -\partial E_0/\partial d$.

Radiative corrections to the Casimir energy have been studied 
for QED in \cite{brw,rsw,bl,xue,zheng,kong1,kong2} and for
scalar field theory in \cite{ford,kay,toms,rein,albu}. Although they are of
no experimental significance in QED (but, possibly within the bag model in
QCD) their correct calculation is a challenge for the ability 
to understand the underlying physical structures and the field 
theoretic methods applied in their study.  
Within standard QED perturbation theory
the calculation of the (leading) $O(\alpha)$ radiative correction to
the Casimir energy (\ref{casienergy}) was performed in \cite{brw} (and
has been confirmed by an independent method in \cite{rsw}). Recently,
the analogous calculation has successfully been completed for the
single sphere geometry in \cite{bl}.  The $O(\alpha)$ radiative
correction originates from the two-loop vacuum diagram shown in Fig.\
1.  Qualitatively, the result for the leading correction $\Delta E_0$
to the ground state energy $E_0$ turns out to be of the order
\begin{equation} 
\label{mc}
{\Delta E_0\over E_0}\sim\alpha{\lambda_{\rm c}\over d},
\end{equation}
where $\alpha$ is the fine structure constant, $\lambda_{\rm c} = 1/m$ 
is the electron Compton wave length ($m$ is the electron mass) 
and $d\gg\lambda_{\rm c}$ 
is the characteristic geometric
length (distance between the mirrors or radius of the sphere, respectively). 
Like in the calculation leading to Eq.\ (\ref{casienergy}), 
it was assumed that for the photon vacuum 
fluctuations conductor boundary conditions 
apply at the ideally reflecting mirror 
surfaces and, in addition, that there are no boundary conditions 
for the spinor field (which would
only contribute geometry-dependent terms which are exponentially 
suppressed for $d\gg\lambda_{\rm c}$ \cite{bhr}). 

Effective field theory methods have widely been used in
various branches of quantum field theory, including QED, in 
recent years (\cite{geor,pich} and references therein). 
They are thought to be able to correctly encapture the relevant
low energy information. As the Casimir effect is a true infrared
phenomenon it appears natural to employ these methods also to
the calculation of Casimir energies. This idea has been pursued
by Kong and Ravndal \cite{kong1,kong2} for two parallel ideally 
reflecting mirrors. In contrast to Eq.\ (\ref{mc}), they find for the 
leading radiative correction $\Delta E_0$ to the Casimir 
energy (\ref{casienergy})
\begin{equation} 
\label{sc}
{\Delta E_0\over E_0}
\sim\alpha^2\left({\lambda_{\rm c}\over d}\right)^4
\end{equation}
which is suppressed relative to Eq.\ (\ref{mc}) by one power of $\alpha$
and three powers of $\lambda_c/d$. Kong and Ravndal observe that Eq.\
(\ref{mc}) cannot be obtained by means of effective field theory
methods (in contrast, Eq.\ (\ref{sc}) can of course be obtained within
full QED as a nonleading correction to Eq.\
(\ref{casienergy})). Although these authors have chosen not to comment
on the problem, looking at \cite{brw,rsw} one quickly recognizes the
reason for this failure.  Effective field theory is always based on a
derivative expansion of the effective action. On the other hand, the
calculation of the correction (\ref{mc}) technically relies on the
consideration of the discontinuity of the (one-loop) photon
polarization operator (see Eq. (18), below) which is part of the QED
effective action and which the Uehling terms discussed by Kong and
Ravndal derive from.  It is rather clear that this discontinuity,
which only starts at the pair production threshold and which,
therefore, has rather to be viewed as a high energy feature, cannot be
seen in any finite order of the derivative expansion of the effective
action (it is a nonperturbative phenomenon with respect to the
derivative expansion). Consequently, effective field theory methods
are unable to reproduce Eq.\ (\ref{mc}). From a formal point of view,
all this is perfectly clear and would not require any further study,
if not the following question with respect to the physical adequacy of
the calculations performed in \cite{brw,rsw,bl} would arise. To what
extent does the result of the formal calculation within full QED which
relies on the discontinuity of the polarization operator encapture
real physics if one takes into account that the frequency threshold
above which the mirrors become transparent is much smaller than the
pair production threshold $2m$?  Could it be that the correction
(\ref{mc}) calculated for ideally reflecting mirrors, although
formally the leading one relative to (\ref{sc}), is suppressed for
realistic mirrors in such a way that the effective field theory result
(\ref{sc}) turns out to be the leading correction to the Casimir
energy (\ref{casienergy}) from a physical point of view? To answer
this question, in the present paper we consider a mathematical model
for penetrable mirrors. We show that the leading radiative correction
to the Casimir energy continues to be of the same order as given in
Eq.\ (\ref{mc}). It depends in a monotonic and powerlike manner on the
parameter describing the threshold at which the mirrors become
transparent. From an effective field theory point of view, 
this result seems to teach an interesting lesson.

The photon propagator in the covariant gauge 
with boundary conditions on a surface $S$ composed of  ideally 
reflecting mirrors
was initially calculated in \cite{brw}, 
a more accessible discussion is given in \cite{bl}. It reads
\begin{equation}
\label{prop3}  
^SD^{\rm c}_{\mu\nu}(x,y) =  
D^{\rm c}_{\mu\nu}(x-y) - \bar{D}_{\mu\nu}(x,y),
\end{equation}
where $D^{\rm c}_{\mu\nu}(x-y)$ is the free-space causal 
propagator and
$\bar{D}_{\mu\nu}(x,y)$ depends on the boundary. 
In plane geometry it can be represented as 
\begin{equation}
\label{prop4}
\bar{D}_{\mu\nu}(x,y)=
\sum _{s=1,2}E_{\mu}^{s}(x)\bar{D}(x,y)E_{\nu}^{s}(y),
\end{equation}
where $E_{\mu}^{s}(x)$ are the two suitably chosen photon 
polarisations which are affected by the
boundaries (see \cite{bl} for details). Here,
$\bar{D}(x,y)$ is the mirror-dependent part of the corresponding
scalar Greens function $^SD^{\rm c}(x,y)$,
which fulfills Dirichlet boundary 
conditions at the mirrors. It appears to be reasonable to apply
the formalism which was used for the derivation of 
the representation (\ref{prop4}) also in the case of penetrable mirrors. 
 From Eq.\ (\ref{prop4}) it is clear that it is sufficient
to concentrate in the following on a massless scalar field which 
fulfills appropriate boundary conditions.

Penetrable mirrors can be modeled by means of delta function
potentials with support at the parallel mirror planes $x_{3}=d_{i}$ 
($i=1,2$)(\cite{bhr,hr}, see \cite{jaek} for a related study). 
For a scalar field of mass $\mu$ the wave equation reads 
\begin{equation} 
\label{eq}
[\Box+\mu^{2}-2a
\sum _{i=1,2}\delta(x_{3}-d_{i})]\varphi(x)=0\ \ .
\end{equation} 
The potential is attractive for $a>0$ (and the spectrum contains bound
states) and it is repulsive for $a<0$. The limit $a\to -\infty$
corresponds to imposing Dirichlet boundary conditions at
$x_3=d_1,d_2$.  Although all subsequent formulas remain valid for
arbitrary values of $a$ we restrict ourselves to $a\le 0$.  The delta
function potential can be reformulated in terms of matching conditions
\begin{eqnarray}
\label{bc}
\varphi(x)_{|_{x_{3}=d_{i}-0}}&=&\varphi(x)_{|_{x_{3}=d_{i}+0}}\ \ ,\\
{\partial\over\partial
  x_{3}}\varphi(x)_{|_{x_{3}=d_{i}-0}}&=&{\partial\over\partial
  x_{3}}\varphi(x)_{|_{x_{3}=d_{i}+0}}\nonumber\\
&&\label{bc1}
\ +\ 2a\ \varphi(x)_{|_{x_{3}=d_{i}}}\ \ .
\end{eqnarray}
The parameter $a$ sets the scale for the (smeared) threshold above
which the mirrors become transparent. For the moment, consider
Eq.\ (\ref{eq}) with just one delta function potential (at $x_3=0$).
The part of its solution depending on $x_3$
\begin{eqnarray}
\varphi(x_3)&=&\left({\rm e}^{ikx_3}+r(k) {\rm e}^{-ikx_3}\right)
\Theta(-x_3)\nonumber\\
&&+t(k) e^{ikx_3}\Theta(x_3)
\end{eqnarray}
has a transmitted and a reflected wave with the coefficients
$r(k)=ia/(k-ia)$, $t=1+r$.  The propagator for a scalar field in the
presence of one delta function potential can be written as
\begin{eqnarray}
\label{prop1d}
&&^{S}D^{\rm c}(x,y)\ =\ \int{{\rm d}^{3}k\over (2\pi)^{3}}
{{\rm e}^{ik_{\beta}(x^{\beta}-y^{\beta})}\over -2i\Gamma}\nonumber\\
&&\times\left\{{\rm e}^{i\Gamma |x_{3}-y_{3}|}-r(\Gamma){\rm
    e}^{i\Gamma(|x_{3}|+|y_{3}|)}\right\}
\end{eqnarray}
with $\Gamma=\sqrt{k_{0}^{2}-k_{1}^{2}-k_{2}^{2}-\mu^{2}+i0}$, $\beta
= 0,1,2$ (cf.\ Eqs.\ (3.10), (3.12) in \cite{bhr}).  When the
reflection coefficient $r(\Gamma)$ approaches zero for 
$\vert\Gamma\vert\gg \vert a\vert $, this propagator turns 
into the free space propagator (in Fourier space). Note, that Eq.\
(\ref{prop1d}) also applies for a general choice of the reflection
coefficient $r(\Gamma)$ which describes dispersive mirrors (entailing
$a\to -i\Gamma r(\Gamma)/[1+r(\Gamma)]$ in the Fourier transformed 
Eqs.\ (\ref{eq}), (\ref{bc1})).

The Casimir energy for two parallel planes with the reflection
coefficient $r(\Gamma)$ corresponding to delta function potentials has
been calculated for scalar and spinor fields in \cite{hr,bhr} and with
a more general choice of $r(\Gamma)$ for a scalar field in
\cite{jaek}. The propagator for a scalar field in the background of
two partly transmitting mirrors modelled by delta function potentials
(cf.\ Eq.\ (\ref{eq})) can be written as $^{S}D^{\rm c}(x,y)=D^{\rm
c}(x-y)-\bar{D}(x,y)$ where $D^{\rm c}(x-y)$ is the free-space causal
propagator and
\begin{eqnarray} 
\label{prop2}
&&\bar{D}(x,y)\ =\ \int{{\rm d}^{3}k\over (2\pi)^{3}} \
{{a\over2}~{\rm e}^{ik_{\beta}(x^{\beta}-y^{\beta})}\over
(\Gamma-ia)^{2}+a^{2}\exp (2i\Gamma d)}\nonumber\\
&&\times\Bigg\{\left(\left(1-{ia\over\Gamma}\right){\rm e}^{i\Gamma
(|x_{3}-d_{1}|+|y_{3}-d_{1}|)}\right.\nonumber\\ && + \left.{ia\over
\Gamma}\ {\rm e}^{i\Gamma(|x_{3}-d_{1}|+|y_{3}-d_{2}|+d)}\right)
+(d_{1}\leftrightarrow d_{2})\Bigg\}
\end{eqnarray} 
is the boundary-dependent part,
$d=\vert d_{2}-d_{1}\vert$
\cite{bhr,hr}\footnote{Eqs.\ 
(3.11), (3.13) in Ref.\ \cite{bhr} contain misprints, the
correct expression is given in Ref.\ \cite{hr}, Eq.\ (7).}.
It is clear, that we need to set $\mu = 0$ in the following.

The $O(\alpha)$ radiative correction to Eq.\ (\ref{casienergy})
has to be derived from the vacuum diagram shown in Fig.\ 1.
The corresponding (divergent) shift in the vacuum energy 
(per unit area of the mirrors, $T V_2$ is the infinite space-time
volume in the directions $\beta = 0,1,2$)
reads
\begin{eqnarray}
\label{de1}
\Delta E_{0}&=&{i\over 2 T V_2}
\int {\rm d}^4x\int {\rm d}^4y\ \left[D^{\rm c}_{\mu\nu}(x-y)\right.\nonumber\\
&&\left. -\bar{D}_{\mu\nu}(x,y)\right]\Pi^{\mu\nu}(x-y)\ \ ,
\end{eqnarray}
where $\Pi_{\mu\nu}(x)=\left[g_{\mu\nu}\Box-\partial_{\mu}\partial_{\nu}\right]
{\Pi}(x^2)$ is the standard one-loop photon polarisation tensor. 
The boundary-independent (divergent) first term connected with the free-space 
propagator $D^{\rm c}_{\mu\nu}(x)$ can be disregarded in the following.
Any effect from the renormalization of the photon polarization tensor 
($\Pi(x^2)\sim {\rm const.}\ \delta^{(4)}(x)$)
can also be disregarded as it also leads to boundary-independent terms
(by virtue of the defining equation of $^SD^{\rm c}_{\mu\nu}(x,y)$).
We can now insert Eqs.\ (\ref{prop4}), (\ref{prop2}) into Eq.\ 
(\ref{de1}). The sum over the Lorentz indices which involves
the sum over the polarisation vectors  
$E_{\mu}^{s}$ can be performed immediately
and simply results in a factor of 2. This corresponds to the known fact 
that in a plane geometry the two photon polarisations have to obey 
the same boundary conditions (this is true for Dirichlet boundary
conditions as well as for the applied penetrable mirrors model).
To proceed further, it is useful to introduce the Fourier 
transform of $\Pi(x^2)$
\begin{equation} 
\label{Pk}
{\Pi}(x^2)=\int{{\rm d}^{4}k\over (2\pi)^{4}}\ 
{\rm e}^{ikx}\ \tilde{\Pi}(k^{2})\ \ .
\end{equation}
Then, in Eq.\ (\ref{de1}) the integrations over 
$x$ and $y$ can be carried out explicitly. 
We arrive at
\begin{equation}
\label{de3}
\Delta E_{0}= 
i\int{{\rm d}^{4}k\over (2\pi)^{4}}\ k^{2}\tilde{\Pi}(k^{2})\tilde{\bar{D}}(k),
\end{equation}
where
\begin{eqnarray}
\label{propfourier}
&&\tilde{\bar{D}}(k)\ =\ -{4\Gamma a\over (k^2)^2}\nonumber\\
&&\times {\left[\Gamma-ia+ia\cos(k_{3}d){\rm e}^{i\Gamma d}\right]
\over
[(\Gamma -ia)^{2}+a^{2}\exp(2i\Gamma d)]}\ \ .
\end{eqnarray}
In deriving Eq.\ (\ref{propfourier}) we have made use of 
the integral ($\Im \Gamma>0$)
\begin{equation} 
\int _{-\infty}^{\infty}{\rm d} x_{3}\  
{\rm e}^{i\Gamma |x_{3}-d|+ik_{3}x_{3}}
={2i\Gamma\over k^{2}}{\rm e}^{ik_{3}d}\ \ .
\end{equation}

It is useful to further transform the representation (\ref{de3}) 
by making use of the analytic
properties of $\tilde{\Pi}(k^{2})$. It has a cut starting at
$k^{2}=4m^{2}$ with the discontinuity ($k^{2} \ge 4m^{2}$)
\begin{eqnarray}
\label{disc}
&&{\rm disc}\tilde{\Pi} (k^{2})\ =\ \tilde{\Pi} (k^{2}+i0) -\tilde{\Pi}
(k^{2}-i0)\nonumber\\
&&=\ -\alpha\ \frac{2i}{3} \sqrt{1-\frac{4m^2}{k^2}}
\left(1+\frac{2m^2}{k^2}\right) \ \ .
\end{eqnarray}
The integration path along the real $k_{3}$-axis can be deformed
upwards into the complex $k_{3}$-plane such a way that 
it encloses the cut (cf.\ Fig.\ 2 on p.\ 045003-8 in \cite{bl}). 
By means of 
\begin{eqnarray} 
&&\int _{-\infty}^{\infty}{\rm d}
k{\tilde{\Pi}(\Gamma^{2}-k^{2})\over \Gamma^{2}-k^{2}}\ [A+B
\cos(kd)]\nonumber\\
&&=-i \int _{2m}^{\infty}{\rm d} q\ {{\rm
    disc}\tilde{\Pi}(q^{2})\over q\sqrt{q^{2}-\Gamma^{2}}}
\ [A+B{\rm e}^{-k_{3}d}]
\end{eqnarray}
($k_{3}=\sqrt{q^{2}-\Gamma^{2}}$)
which entails a change of variable we obtain
\begin{eqnarray}
&&\Delta E_{0}\ =\ -4a\int{{\rm d}^{3}k\over (2\pi)^{3}}\
\int _{2m}^{\infty}{{\rm d} q\over 2\pi}\ {{\rm
    disc}\tilde{\Pi}(q^{2})\over q\sqrt{q^{2}-\Gamma^{2}}}\nonumber\\
&&\times{\Gamma(\Gamma -ia+ia {\rm e}^{-k_{3}d}{\rm e}^{i\Gamma d})\over 
(\Gamma-ia)^{2}+a^{2}\exp (2i\Gamma d)}\ \ .
\end{eqnarray}
Furthermore, the Wick rotation $k_{0}\to ik_{4}$,
$\Gamma\to i\gamma=i\sqrt{k_{4}^{2}+k_{1}^{2}+k_{2}^{2}}$
can be performed. The resulting expression for $\Delta E_0$ 
still contains ultraviolet divergencies which, however, do 
not depend on the distance between the mirrors (these divergencies
arise from immersing single mirrors into the vacuum). 
After subtracting these divergent terms the finite, distance-dependent part
of $\Delta E_{0}$ finally reads
\begin{eqnarray}
\label{finenergy}
&&\Delta E_{0}\ =\ -4ia\int{{\rm d}^{3}k_{\rm E}\over (2\pi)^{3}}
\int _{2m}^{\infty}{{\rm d} 
q\over 2\pi}{{\rm
    disc}\tilde{\Pi}(q^{2})\over q\sqrt{q^{2}+\gamma^{2}}}\nonumber\\
&&\times {\gamma a 
{\rm e}^{-\gamma d}[a{\rm e}^{-\gamma d}+(\gamma -a){\rm e}^{-k_3 d}]
 \over 
(\gamma -a)[(\gamma-a)^{2}-a^{2}\exp(-2\gamma d)]}\ \ .
\end{eqnarray}
In the limit $a\to -\infty$, which can be performed in the
integrand, the expression for impenetrable mirrors, with Dirichlet boundary
conditions, is reobtained. 

We are interested in the leading term of Eq.\ (\ref{finenergy})
for $d/\lambda_{\rm c}= md\gg 1$.  
Then, in Eq.\ (\ref{finenergy}) the contribution containing
$\exp(-k_{3}d)$ [$\le \exp(-2md)$] in the nominator can be neglected at
leading order.  Also, in the denominator we can approximate
$\sqrt{q^{2}+\gamma^{2}}$ by $q$. So the integral related to the fermion loop decouples
in this approximation and with
\begin{equation}
\label{cutint}
\int _{2m}^{\infty}{\rm d} q
{{\rm disc}\tilde{\Pi}
\left(q^{2}\right) \over q^{2}}= 
-\frac{3i\pi}{32}{\alpha\over m}
\end{equation}
we obtain to leading order in $1/md$
\begin{eqnarray}
\label{erg1}
&&\Delta E_{0}\ =\ - {3\over 32\pi^2}{\alpha a^{3}\over m d}\ 
\int^\infty_0{\rm d}\gamma
\nonumber\\  
&&\times {\gamma^3 {\rm e}^{-2\gamma}\over 
(\gamma -ad)[(\gamma -ad)^{2}-(ad)^{2}\exp(-2\gamma)]}\ \ .
\end{eqnarray}
It looks structurally quite similar to the 
analogous expression (4.4) in \cite{bhr} for the
Casimir energy $E_0$ itself.
As a function of $a$ it is monotonic (for $a<0$). The limiting cases are
\begin{equation}
\Delta E_0 \stackrel{ad\to 0}{=} {\alpha\over md}{3\over 64\pi^2}(-ad)^3
+\ldots
\end{equation}
and 
\begin{eqnarray}
\label{fin1}
&&\Delta E_0 
\stackrel{ad\to -\infty}{=} 
{\alpha\over md}{\pi^2\over 2560}{1\over d^3}\left(1-{\sigma\over -ad}+\ldots\right)\ \ 
\end{eqnarray}
with $\sigma=4(1+45\zeta (5)/\pi^4)\sim 5.92$.

Using the expression for the Casimir energy $E_0$ obtained in
\cite{bhr}\footnote{In Eq.\ (4.7) in \cite{bhr} a factor of 1/2 is missing.} 
the relative weight of the radiative correction $\Delta E_0$, 
i.e., the ratio (\ref{mc}), can be shown to be a monotonic function of $a$ 
with the limiting cases
\begin{equation}
{\Delta E_0\over E_0} \stackrel{ad\to 0}{=}
{\alpha\over md}{3\over 16}\ {a  d} +\ldots
\end{equation}
and 
\begin{eqnarray}
\label{fin2}
&&{\Delta E_0\over E_0}
\stackrel{ad\to -\infty}{=}
{\alpha\over md}{-9\over 32}
\left(1-{\sigma-3\over -ad}+\ldots\right)\ \ .
\end{eqnarray}
 From Eqs.\ (\ref{fin1}), (\ref{fin2}), which apply in the physically
interesting domain $-ad\gg1$, we deduce the result stated above, that the
$O(\alpha)$ radiative correction to the Casimir energy experiences a
powerlike modification for penetrable mirrors. 

Acknowledgements: We would like to thank G.\ Barton, K.\ Kirsten, J.\
Lindig, and D.\ V.\ Vassilevich for reading a draft version of the
paper and comments on it.

\begin{figure}[ht]
\unitlength1cm
\begin{picture}(14,2.5)
\label{fd}
\put(3,1){\epsffile{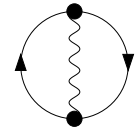}}
\end{picture}
\caption{Two-loop vacuum diagram which the leading $O(\alpha)$
correction to the Casimir energy originates from}
\end{figure}


\begin{thebibliography}{99}

\bibitem{vacu}
P.\ W.\ Milonni, {\it The Quantum Vacuum -- An Introduction to 
Quantum Electrodynamics} (Academic Press, Boston, 1994).

\bibitem{casi1}
H.\ B.\ G.\ Casimir, Proc.\ K.\ Ned.\ Akad.\ Wet.\ {\bf 51}, 793 (1948).

\bibitem{casi2}
V.\ M.\ Mostepanenko and N.\ N.\ Trunov, {\it The Casimir Effect 
and its Applications} (Oxford University Press, Oxford, 1997).

\bibitem{lamo}
S.\ K.\ Lamoreaux, Phys.\ Rev.\ Lett.\ {\bf 78}, 5 (1997).

\bibitem{mohi}
U.\ Mohideen and Anushree Roy, {\it A Precision Measurement of 
the Casimir Force from 0.1 to 0.9 Microns}, physics/9805038.

\bibitem{brw}
M.\ Bordag, D.\ Robaschik and E.\ Wieczorek, Ann.\ Phys.\ (N.Y.)
{\bf 164}, 192 (1985).

\bibitem{rsw}
D.\ Robaschik, K.\ Scharnhorst, and E.\ Wieczorek, Ann.\ Phys.\ (N.Y.)
{\bf 174}, 401 (1987).

\bibitem{bl}
M.\ Bordag and J.\ Lindig, Phys.\ Rev.\ D {\bf 58}, 045003 (1998). 

\bibitem{xue}
She-Sheng Xue, Commun.\ Theor.\ Phys.\ (Wuhan) {\bf 11}, 243 (1989).

\bibitem{zheng}
Tai-Yu Zheng and She-Sheng Xue, Chin.\ Sci.\ Bull.\ {\bf 38}, 631
(1993).

\bibitem{kong1}
X.\ Kong and F.\ Ravndal, Phys.\ Rev.\ Lett.\ {\bf 79}, 545 (1997).

\bibitem{kong2}
X.\ Kong and F.\ Ravndal, {\it Quantum Corrections to the QED
Vacuum Energy}, hep-ph/9803216, Nucl.\ Phys.\ B to appear.

\bibitem{ford}
L.\ H.\ Ford, Proc.\ R.\ Soc.\ London, Ser.\ A {\bf 368}, 305 (1979).

\bibitem{kay}
B.\ S.\ Kay, Phys.\ Rev.\ D {\bf 20}, 3052 (1979).

\bibitem{toms}
D.\ J.\ Toms, Rev.\ D {\bf 21}, 2805 (1980).

\bibitem{rein}
K.\ Langfeld, F.\ Schm\"user, and H.\ Reinhardt, Phys.\
Rev.\ D {\bf 51}, 765 (1995).

\bibitem{albu}
L.\ C.\ de Albuquerque, Phys.\ Rev.\ D {\bf 55}, 7754 (1997).

\bibitem{bhr}
M.\ Bordag, D.\ Hennig, and D.\ Robaschik, J.\ Phys.\ A
{\bf 25}, 4483 (1992).

\bibitem{geor}
H.\ Georgi, Annu.\ Rev.\ Nucl.\ Part.\ Sci.\ {\bf 43}, 209 (1993).

\bibitem{pich}
A.\ Pich, {\it Effective Field Theory}, Lecture at the Les Houches
Summer School in Theoretical Physics, Session 68: Probing the 
Standard Model of Particle Interactions, Les Houches, France,
July 28 - September 5, 1997,
University of Valencia preprint FTUV/98-46, IFIC/98-47,
hep-ph/9806303.

\bibitem{hr}
D.\ Hennig and D.\ Robaschik, Phys.\ Lett.\ A {\bf 151}, 209 (1990).

\bibitem{jaek}
M.\ T.\ Jaekel and S.\ Reynaud, J.\ Phys.\ I France {\bf 1}, 1395 (1991).


\end{thebibliography}
\end{document}